# Ising model: secondary phase transition

You-gang Feng

College of Science, Guizhou University, Huaxi, Guiyang, 550025 China
E-mail: ygfeng45@yahoo.com.cn

**Abstract** Lttice-spin phonons are considered, which make the heat capacity at the critical temperature satisfy experimental observations better. There is a BEC phase transition in an Ising model attributable to the lattice-spin phonons. We proved that the spin-wave theory only is available after BEC transition, and the magnons have the same characteristics as the lattice-spin phonons', resulting from quantum effect. Energy-level overlap effect at ultralow temperature is found. A prediction of BEC phase transition in a crystal is put forward as our theory generalization.

## 1. Introduction

We have been succeeded in initially revealing the nature of ordered Ising model at the critical temperature $T_c$ in that there exist elementary excitations, which may cause nonsingular heat capacities [1]. The system is recognized as a new normal one, the block-spin phonons and the sub-block spin phonons do show the system characteristics. In this paper we will explore further for the properties of an ordered Ising model. Experimental observations have found that in the region of temperature $T < T_c$ near $T_c$ the heat capacity approaches in very high power-exponent law to its maximum at $T_c$ but singularity. The heat capacities, however, attributed to block-spin phonons and sub-block spin phonons cannot enough achieve the expected phenomenon. We should consider this case with a more suitable theory, and study on fine structures of elementary excitations. The formation of blocks is a result of the locally strong correlation of lattice spins. The correlation has wave-motion property to cause elementary excitation in blocks, which quanta are called lattice-spin phonons. The heat capacity stemming from the lattice-spin phonons should be considered. In section 2 of this paper, we will see that sub-block spin phonons and block-spin phonons may disappear with vanishing of fluctuation, only the lattice-spin phonons exist when temperature decreases. In addition, we notice that the lattice-spin phonons can be regarded as ideal Bose gas, its special property at low temperature should be reviewed. Since Bose-Einstein condensation (BEC) in dilute atomic gases has produced by experimental groups in 1995 [2-5], the experimental observations and relevant theoretical explanations for BEC phenomena in diverse physical systems have successively been reported [6-14], among them the interesting issues for us are that refer to BEC of magnons [8-12]. The similarity between magnons and lattice-spin phonons enable us to believe that an Ising model, as an example of ferromagnet, may experience a BEC phase transition at ultralow temperature, which will determine a characteristic of magnetic saturation. In section3, we discuss the $T^{2/3}$ law in BEC process, and link it to quantum effect. Comparison of lattice-spin phonons with magnons has demonstrated that spin-wave theory is only available after BEC phase transition, having the same properties as the lattice-spin phonons'. Energy-level overlap effect at ultralow temperature is discussed. There is a change in mechanism of heat capacity, there is a continuous phase transition. Finally, a prediction of BEC phase transition in a crystal is given as generalizing our research on the low-temperature behaviour of Ising models.

## 2. Theory
*2.1 Lattice-spin phonons in blocks*
The characteristic of correlation for inner lattice spins in a block can be considered from two hands. On the one hand, the conservation of inner product of a block spin shows that the correlation obeys rotation symmetry, behaving as a harmonic motion. On the other hand, according to the hierarchical property of blocks the inner lattice spins of an r-order block on the r-th hierarchy are originally the (r-1)-order blocks on the (r-1)-th hierarchy [1], the property of these lattices is similar to the one of the (r-1)-order blocks due to the self-similarity. This means that the lattice spin correlation, like the block spins', demonstrates wave feature, which relevant quanta are called lattice-spin phonons. The inside space of a block amounts to a *D*-dimensional hypercube with side *n* and lattices $P = n^D$ equal to the number of solid-state physics primitive cells [15]. A periodic boundary condition is introduced, and the components of wave vector ***k*** are

$$k_{x_j} = 0; \pm 2\pi/n; \pm 4\pi/n; \cdots; P\pi/n. \tag{1}$$

where $x_j$ is the coordinate of the *j*-th axis, and $j = 1,2,\cdots, D$. Each volume element $(2\pi/n)^D$ contains one *k*, i.e. there are allowable values of $(n/2\pi)^D$ per unit volume in the ***k*** space. The product of $(n/2\pi)^D$ by a sphere volume of radium *k* gives a total number of models, and the related wave-vector magnitudes are smaller than *k*. For a *D*-dimensional hypersphere its volume is $Lk^D$, where *L* is a constant (see Appendix A). We then get a relation for each polarization: $P = (n/2\pi)^D \cdot Lk^D$. A dispersion relation is given by $\omega = vk$, where $\omega$ is angular frequency and *v* is sound velocity. Since there are infinite blocks the total number of inner lattice spins are also infinite, therefore we can describe the lattice-spin phonons in statistical law. A state density for each polarization is

$$D(\omega) = dP/d\omega = (n/2\pi v)^D \cdot LD\omega^{D-1} \tag{2}$$

The maximum of *k* determines a Debye's cutoff frequency $\omega_D$: $\omega_D = 2\pi v \cdot L^{-1/D}$.

The inside space of a block is $D$ dimensions, being equal to independent degrees of freedom for a lattice-spin phonon. The thermal energy contributed by the lattice-spin phonons in each polarization is written by

$$U = \int_0^{\omega_D} D(\omega) <n(\omega)> \hbar\omega \, d\omega \qquad (3)$$

Where $<n(\omega)>$ is the mean quantum number at temperature $T$ in the energy levels with frequency $\omega$, and $<n(\omega)> = [\exp(\hbar\omega/k_B T) - 1]^{-1}$, $\hbar$ is Plank constant, $k_B$ Boltzmann constant. For a homogeneous system with D polarization directions the total energy equals

$$<E> = DU = \frac{LPD^2}{(2\pi v)^D} \int_0^{\omega_D} \frac{\hbar\omega^D}{\exp[(\hbar\omega)/(k_B T)] - 1} \, d\omega \qquad (4)$$

Therefore, we get the heat capacity of lattice-spin phonons for a constant number $P$ of lattice spins:

$$C_v = (\partial <E>/\partial T)_P = LPD^2 k_B \left(\frac{k_B T}{2\pi v \hbar}\right)^D \int_0^{x_D} \frac{x^{D+1} e^x}{(e^x - 1)^2} \, dx \qquad (5)$$

where $x = \hbar\omega/(k_B T)$, $x_D = \hbar\omega_D/(k_B T)$. Equation (5) is suitable to all inner lattice spins of block spins and of sub-block spins. For example, in a cubic lattice spin system a $D$-type spin has a $D$-dimensional inside space, a $D^4$-type spin $D^4$-dimensional inside space. The heat capacity at $T_c$ attributed to all types of phonons takes

$$C_v = a_1 T_c^2 + a_2 T_c^3 + b_1 T_c^D + b_2 T_c^{D^4} \qquad (6)$$

where $a_1$, $a_2$, $b_1$, and $b_2$ are constants, $T_c^2$ and $T_c^3$ are related to the sub-block spin phonons and the block-spin phonons, respectively. $D_c^D$ and $T_c^{D^4}$ result from the lattice-spin phonons in $D$-type spins and in $D^4$-type spins, respectively. By reference [15], the numerical calculation indicates that a small fractal dimension with integer side in close vicinity to the critical point is about $D \cong 2.4781$, and $D^4 = 37.7116$. Inserting these data into equation (6) gives a heat capacity with very high power exponent:

$$C_v = a_1 T_c^2 + a_2 T_c^3 + b_1 T_c^{2.4781} + b_2 T_c^{37.7116} \qquad (7)$$

Other fractal dimensions that are greater than the above one may also occur due to the fluctuation. It is expected that more high power exponents will appear in keeping with the experimental observations.

The fractal structures of blocks will not disappear suddenly when temperature decreases from the critical temperature, the heat capacity of a system depends on a summation of Debye's models for all kinds of spin phonons such as block-spin phonons and lattice-spin phonons of blocks. A system magnetization only is determined by the block-spin phonons since spin is a vector. Let $N_b$ denote a total number of block-spin phonons, $N_b^{ex}$ denote a number of block-spin phonons in excitation states and $N_b^{ex} = \int D_b(\omega) <n(\omega)> \mathrm{d}\omega$, $D_b(\omega) = V\omega^2/(2\pi^2 v^3)$ is the state density of block-spin phonons, $V$ is the system volume. Let $M(T)$ be a magnetization at temperature $T$, $M(0)$ be a magnetization at absolute zero. A relative magnetization is $M(T)/M(0)$, and $M(T)/M(0) = 1 - N_b^{ex}/N_b$, we then have

$$M(T)/M(0) = 1 - \frac{V}{2\pi^2 v^3 N_b}\left(\frac{k_B T}{\hbar}\right)^3 \int_0^{x_D} \frac{x^2}{e^x - 1}\mathrm{d}x \qquad (8)$$

where the Debye frequency of block-spin phonons is given by $\omega_D^3 = 6\pi v^3 N_b/V$. The phonon density depends on the block size, if the size changes the relative magnetization per unit volume will change too, even though the temperature is constant.

*2.2 Lattice-spin phonons of system*
The formation of a block results from the locally strong correlation of lattice spins, which leads to a deviation in spin states between the lattice spins outside blocks and the lattice spins inside blocks The deviation is just the fluctuation. The system is capable of eliminating the fluctuation at thermal equilibrium without foreign field as the fluctuation-dissipation theorem [16], such that the extent of correlation for lattice spins expands so rapidly when the temperature slowly decreases that the whole system becomes a block with infinite side, and there is not any other block in the temperature region not far from the critical temperature. With the same reason as introduction of block-spin phonons we can also introduce lattice-spin phonons with a spin parameter *q*. The heat capacity of system and the relative magnetization are described by Debye model of lattice-spin phonons of the system. In a low-temperature region, when $x_D \gg 1$, the heat capacity $C_v$ becomes [17]

$$C_v \cong 234 N_L k_B (T/\theta)^3 \qquad (9)$$

where $\theta = \hbar\omega_D/k_B$, called Debye temperature, $N_L$ denotes a total number of

lattice-spin phonons. Subsequently, the relative magnetization is

$$M(T)/M(0) \propto 1 - \alpha_L T^3 \tag{10}$$

where $\alpha_L$ is a constant, $M(T)/M(0)$ is a convex function of $T$.

*2.3 BEC phase transition*
Since lattice-spin phonons are assigned as ideal gases, when temperature decreases to a turning point the BEC of the gases as strongly degenerate condition become so significant that the system properties will be changed dramatically different from ones above the point. We call the turning point a critical temperature $T_{BE}$ of BEC [18], which is

$$T_{BE} = \frac{h^2}{2\pi m k_B} \cdot [\frac{N}{V\xi(3/2)}]^{3/2} \tag{11}$$

where $h = 2\pi\hbar$, $m$ is effective mass of a phonon, $\xi(3/2)$ is a Rieman zeta function, and $\xi(3/2) \cong 2.612$. The ratio of the number $N_L^{ex}$ of phonons in excitation state to the total number $N_L$ of phonons is given by

$$N_L^{ex}/N_L = (T/T_{BE})^{3/2}, \qquad (T < T_{BE}) \tag{12}$$

The relevant magnetization is

$$M(T)/M(0) = [1 - (T/T_{BE})^{3/2}] \tag{13}$$

We then get the heat capacity of the system

$$C_v \propto (T/T_{BE})^{3/2} \tag{14}$$

Equation (13) shows that the function of $M(T)/M(0)$ versus $T$ is a convex curve, and its increases slowly as temperature decreases to demonstrate a characteristic of magnetic saturation. We call uniformly the laws exhibited by equations (12)-(14) a $T^{3/2}$ law. Comparison of equations (13) and (14) with equations (9) and (10) indicates that the turning point $T_{BE}$ is another critical temperature of continuous phase transition for an Ising model.

**3. Discussion**
*3.1 Quantum effect and $T^{3/2}$ law*
The fact that our $T^{3/2}$ law of heat capacity and relative magnetization accords with the conclusion of spin-wave theory reveals a connection between the two theories. The spin-wave theory put forward by Dyson starts with the assumption that quantum collision effect can produce the spatial-temporal spin oscillations, which states propagate in a wave manner in a 3-dimensional ferromagnetism such as a cubic lattice spin system [19-21]. The BEC of lattice-spin phonons also arise from a quantum effect, the wave packets of phonons overlap so tightly that the mean interparticle separation becomes comparable to the particle's thermal wavelength, such that the quantum attraction leads to phonon condensation in ground state. We may owe the $T^{3/2}$ law to the quantum effect, and the spin-wave theory is only operable to the temperature below $T_{BE}$. Therefore, the two kinds of quasiparticles are the same thing.

In fact M. Matsubara and H. Matsuda first called attention to the magnon's characteristics by pointing out that a quantum spin system is equivalent to an interacting Bose gas [22]. The spin-wave theory, however, cannot predict the BEC phase transition.

To obtain a clear insight into the relation between the $T^{3/2}$ law and the quantum effect we introduce a revised Bloch's spin-wave model [23], and apply it to a simply cubic lattice spin system in the following discussion. At first, we don't confine the spin orientation to a particular direction, and the components in *x*, *y*, and *z* axes are nonzero. There are only the nearest neighbor interactions. A Hamiltonian of exchanging action of these spins takes

$$H = -2J \sum_R \boldsymbol{S}_R \cdot \boldsymbol{S}_{R+\tau} \tag{15}$$

Where *J* is an exchange integral, $\boldsymbol{S}_R$ is a spin angular momentum with a position vector $\boldsymbol{R}$, which adjacent one has position vector $\boldsymbol{R}+\tau$. The relevant Poisson bracket is

$$d\boldsymbol{S}_R/dt = -\frac{i}{\hbar}[\boldsymbol{S}_R, H] \tag{16}$$

where the letter *i* denotes a pure imaginary, *t* is time. From equation (16) we get a component equation

$$d S_R^j = -\frac{i}{\hbar}[S_R^j H - H S_R^j] \tag{17}$$

where $j = x, y, z$. The spin matrices are represented as

$$S^x = \frac{\hbar}{2}\begin{pmatrix} 0 & 1 \\ 1 & 0 \end{pmatrix} \; ; \; S^y = \frac{\hbar}{2}\begin{pmatrix} 0 & -i \\ i & 0 \end{pmatrix} \; ; \; S^z = \frac{\hbar}{2}\begin{pmatrix} 1 & 0 \\ 0 & -1 \end{pmatrix} \tag{18}$$

We then get

$$S^x S^y - S^y S^x = i\hbar S^z ; \; S^y S^z - S^z S^y = i\hbar S^x ; \; S^z S^x - S^x S^z = i\hbar S^y \tag{19}$$

Using equation (19), we have

$$d S_R^x / dt = 2J(S_R^y \sum_\tau S_{R+\tau}^z - S_R^z \sum_\tau S_{R+\tau}^y) \tag{20}$$

By the same treating, we can get $d S_R^y / dt$ and $d S_R^z / dt$ equations similar to equation

(20). According to Bloch's suggestion there is only one spin to be antiparallel to all other spins parallel to z-axis in the immediate of absolute zero, resulting in a small increasing energy. The increment of energy amounts to that a negative infinitely small spin is additionally apportioned to each of lattice spins, and all spins are still visualized as parallel to one another along z-axis, such treatment is different from Bloch's. The approach makes us identify the system with an Ising model, meeting the requirements of a classical model. In the quantum mechanics sense, however, the zero components of spins in x and y axes can be considered statistical mean values for $S^x$ and $S^y$, respectively. We think that an electron is responsible for a lattice spin, its quantum effect cannot be neglected at such low temperature, namely, we should make much account of the quantum fluctuation in spins about their mean values. Uncertainty principle that is markedly predominant gives all lattice spins additional increments of negative or positive values in x, y, and z axes, such that we obtain: $0 < S^x, S^y \ll S$, $S^z \cong S$, where $S$ is the magnitude of a lattice spin at high temperature, it is a constant. Therefore, from equation (20) we have

$$d\,SR_R^x / dt = 2JS(ZS_R^y - \sum_\tau S_{R+\tau}^y) \tag{21.1}$$

$$d\,S_R^y / dt = -2JS(ZS_R^x - \sum_\tau S_{R+\tau}^x) \tag{21.2}$$

$$d\,S_R^z / dt \cong 0 \tag{21.3}$$

where the coordinate number is Z=6 for a simple cubic system. The solution of equation (21) is

$$S_R^x = F\exp[i(\mathbf{kR} - \omega t)]; \qquad S_R^y = G\exp[i(\mathbf{kR} - \omega t)] \tag{22}$$

Inserting equation (22) into equation (21) yields

$$-i\omega F = 2JS[6 - \sum_\tau \exp(i\mathbf{k}\tau)]G; \quad i\omega G = 2JS[6 - \sum_\tau \exp(i\mathbf{k}\tau)]F \tag{23}$$

The equations have a solution for F and G if the determination of the coefficients is equal to zero, from which we get

$$\hbar\omega = 2JS\hbar[6 - \sum_\tau \exp(i\mathbf{k}\tau)] \tag{24}$$

Long-wavelength limit condition holds at low temperature. We expand the exponent term of equation (24) in series about k, and keep quadratic term, and get

$$\omega = (2JSa^2)k^2 \tag{25}$$

where the lattice constant is $a$. Equation (25) shows a typical feature of the dispersion relation of $\omega$ and k for magnons. Using Bose-Einstein statistics and equation (24), we can get the same results for $C_v$ and $M(T)/M(0)$ as the previous ones. That the theories mentioned above get the same result proves again that quantum effect plays a role of the utmost important for the $T^{3/2}$ law.

The following facts give two other pieces of evidence for our reasoning. In a spin-gap magnetic compound $TlCuCl_3$ a BEC of magnons is found upon application of a magnetic field, $T^{3/2}$ law occurs at temperature below $T_{BE}$, consisting with theoretical prediction [9]. The same law is also discovered in a system of $NiCl_2$-$4SC(NH_2)_2$, which transition under a critical magnetic intensity can be theoretically interpreted as a phenomenon of BEC for magnons [8], and the theory different from that is used in the first case.

*3.2 Energy-level overlap effect*

An energy level $E_n$ of a lattice-spin phonon with quantum number *n* is expressed as $E_n = (1/2 + n)\hbar\omega$ analogous to a quantum-harmonic oscillator's. For a harmonic oscillator system $\omega$ is unique; for an Ising model, however, it is only an element of a set of frequencies. This means that there is an energy-level spectrum of ground states ranging from zero frequency to a maximum $\omega_D$. Since there are infinite lattice spins the frequency values vary continuously such that it is possible that the energy-levels of ground states overlap partially ones of excitation states, and an energy-level of ground state may equal an energy-level of excitation state, if the former frequency is high enough to the latter one. This effect is equivalent to that the Debye frequency $\omega_D$ becomes higher to increase the number of lattice-spin phonons in excitation state without losing energy, and lowers its frequency. Such crucial transfer occurs only in the vicinity of absolute zero, while the asymptotic process of magnetic saturation will slow down.

Such effect will also occur in a crystal. The early experimental observations found that the Debye frequency anomalously increases when temperature decreases to absolute zero, regardless of the effect [24-25].

*3.3 Prediction about BEC phase transition in a crystal*

Since the phonons in a crystal are very much the same as the lattice-spin phonons in an Ising model, an effect of BEC for phonons in a crystal is expected to exhibit at the temperature near absolute zero. Therefore, the $T^3$ law is no more than behaviour of phonons at low temperature before the BEC phase transition. We believe that modern advanced experimental technique will help us demonstrate evidence of this potential phenomenon heralding a $T^{3/2}$ law for heat capacity.

**4. Conclusion**

There are three characteristic regions of temperature for an ordered 3-dimensional

Ising model. The first one spans a range from absolute zero to the second critical temperature $T_{BE}$ wherein $T^{3/2}$ law is attributable to BEC of lattice-spin phonons. The third refers to the scope of temperature being near and involving the first critical temperature $T_c$, where there are fractal structures of blocks, and the block-spin phonons, the sub-block spin phonons, and the lattice-spin phonons are responsible for the heat capacity by their Debye models. $M(T)/M(0)$ is only determined by the block-spin phonons. The second region is in between the above two regions, in which only lattice-spin phonons play role due to there is no block or sub-block. We here call uniformly sub-block spin phonons, block-spin phonons, and lattice-spin phonons and their relevant theories spin phonons and spin-phonon theory, respectively. Clearly, the application of fractal theory and spin-phonon theory extends the extent of Ising model research, the model not only shows the well-known properties in the range from high temperature to the first critical temperature $T_c$, but also exhibits many other distinct features from the temperature near absolute zero to the $T_c$, involving the existence of the second critical temperature $T_{BE}$. This versatile model is fairly favorable for us to understand the mystery nature of matter.

## Appendix A

Suppose there are two hypercubes with volumes $L_1^D$ and $L_2^D$ in a D-dimensional space, a hypersphere of volume $\Omega$ is their inscribed sphere and circumscribed sphere, respectively, i.e. $L_1^D < \Omega < L_2^D$. Clearly, there exists a hypercube of side $L_3$, and it satisfies a condition: $L_1 < L_3 < L_2$, which makes an identity hold: $L_3^D = \Omega$. If the hypersphere has a radium $R$, we certainly have $\Omega = LR^D$, and $L = (L_3/R)^D$, since topology tells us that a sphere and a cube are homoeomorphism only if both have the same dimensionality. When $D$ is a natural number, $L$ is a Gamma function with $D$ as its variable [26]. For the fractal dimension $L$ is a constant related to $D$.

**References**
[1] You-gang Feng 2011 *arxiv:1111.2233*
[2] Davis K.B., M.O.Mewes, M.R.Andreus, N.J.van Druten, D.S.Durfee, D.M.Kurn, and W.Ketterle 1995 *Phys.Rev.Lett.* **75** 3469
[3] Anderson, M.H., J.R.Matthew, C.E.Wieman, and E.A.Cornell 1995 *Science* **269** 198

[4] Newbury N.R., C.J.Myatt and C.E. Wieman 1995 *Phys.Rev. A* **51** R2680
[5] Wolfgang Petrich, Michael H.Anderso, Jason R.Ensher, And Eric A.Cornell 1995 *Phys.Rev.Lett.* **74** 3352
[6] M.Erhard, H.Schmaljoham, J.Kronjager, K.Bongs, and K.Sengtock 2004 *Phys.Rev. A* **70** 031602
[7] Axel Griesmaier, Jorg Werner, Sven Hensler, Jurgen Stubler, and Tilman Pfau 2005 *Phys.Rev.Lett.* **94** 160401
[8] D.Reyes, A.Paduan-Filho, and M.A.Continentino 2008 *Phys.Rev. B* **77** 052405
[9] T.Nikuni, M.Oshikawa, A.Oosawa and H.Tanaka 2000 *Phys.Rev.Lett.* **84** 5868
[10] M.Jaime, V.F.Corra, N.Harrison, C.D.Batista, N.Kawashima, Y.Kazuma, G.A.Jorge, R.Stern, I.Heinma, S.A.Zvyagin, Y.Sasago, and K.Uchinokura 2004 *Phys.Rev.Lett.* **93** 087203
[11] J.Sirker, A.Weisse and O.P.Sushkov 2004 *EPL.* **86** 275
[12] A.V.Syromyanikov 2007 *Phys.Rev. B* **75** 134421
[13] Dagim Tilakum, R.A.Duine, and A.H.MacDonald 2011 *Phys.Rev. A* **84** 033622
[14] K.Mikelson and J.K.Freericks 2011 *Phys.Rev A* **83** 043609
[15] You-gang Feng 2010 *arxiv:1007.1503*
[16] Kubo R. 1996 *Rep. Prog. Phys.* **29** 255-284
[17] Charles Kittel 1996 *Introduction to solid state physics* 7[th]. Edit. (New York, John Wiley) p124
[18] R.K.Pathria 2001 *Statistical mechanics* 2[nd]. Edit. (Singapore, Elsevier) p157-168
[19] T.Holstein, H.Primakoff 1940 *Phys Rev.* **58** 1098
[20] F.Dyson 1956 *Phys.Rev.* **102** 1217
[21] C.Herrig and C.Kittel 1951 *Phys.Rev.* **81** 869
[22] T.Matsubara and H.Matsuda 1956 *Prog.Theor.Phys.* **16** 569
[23] F.Bloch 1930 *Z.Physik* **61** 206; 1932 *Z.Physik* **74** 295
[24] M.Blackman 1935 *Proc.Roy.Soc.* **A148** 365, 385; **A149** 117, 126
[25] E.W.Kellermann 1941 *Proc.Roy.Soc.* **A178** 17
[26] Huber Greg 1982 *Am.Math.Monthly* **89** 301